\documentclass[aps,pra,groupedaddress,floatfix,nofootinbib]{revtex4}

\begin{document}

\title{On the homotopy perturbation method for Boussinesq--like equations}
\author{Francisco M. Fern\'{a}ndez}

\affiliation{INIFTA (UNLP, CCT La Plata-CONICET), Divisi\'on Qu\'imica
Te\'orica Blvd. 113 y 64 S/N, Sucursal 4, Casilla de Correo 16, 1900 La
Plata, Argentina}

\begin{abstract}
We comment on some analytical solutions to a class of Boussinesq--like
equations derived recently by means of the homotopy perturbation method
(HPM). We show that one may obtain exactly the same result by means of the
Taylor series in the time variable. We derive more general results by means
of travelling waves and argue that a curious superposition principle may not
be of any mathematical or physical significance.
\end{abstract}

\maketitle

In a recent paper Y\i ld\i r\i m\cite{Y09} solved a kind of generalized
Boussinesq--like equations by means of the homotopy perturbation method
(HPM). The purpose of this comment is the discussion of the results derived
by that author and the conclusions drawn from them.

Y\i ld\i r\i m\cite{Y09} investigated $B(m,n)$ equations of the form
\begin{equation}
u_{tt}=(u^{m})_{xx}-(u^{n})_{xxxx},\,m,n>1  \label{eq:B(m,n)_0}
\end{equation}
were the subscripts denote derivatives of the solution $u(x,t)$ with respect
to its arguments. However, in the examples Y\i ld\i r\i m changed his mind
and considered particular cases of
\begin{equation}
u_{tt}=-(u^{m})_{xx}+(u^{n})_{xxxx}  \label{eq:B(m,n)}
\end{equation}
Both equations exhibit completely different solutions and we mainly restrict
our discussion to Eq.~(\ref{eq:B(m,n)}) in what follows.

Y\i ld\i r\i m applied HPM which looks like an ordinary perturbation theory
with a fancy name. After introducing a perturbation parameter $p$, Y\i ld\i
r\i m obtained a perturbation series of the form
\begin{equation}
u(x,t)=\sum_{j=0}^{\infty }v_{j}(x,t)p^{j}  \label{eq:u_p_series}
\end{equation}
and stated that its convergence was proved by He\cite{H04}. However, that
reference does not contain any convergence proof. As far as we know, the
proof of convergence was suggested to He by an anonymous referee of an
earlier paper\cite{H99} but to our knowledge no HPM user has ever verified
if his particular implementation of the approach already met those
convergence criteria.

After substituting $p=1$ in the resulting perturbation series (\ref
{eq:u_p_series}) Y\i ld\i r\i m realized that it was merely the Taylor
series about $t=0$:
\begin{equation}
u(x,t)=\sum_{j=0}^{\infty }f_{j}(x)t^{j}  \label{eq:u_t_series}
\end{equation}
where $f_{0}(x)=u(x,0)$ and $f_{1}(x)=u_{t}(x,0)$. By inspection of the
Taylor series Y\i ld\i r\i m\cite{Y09} identified some exact analytical
solutions to the nonlinear partial differential equations. It is worth
noticing that Y\i ld\i r\i m chose the particular initial conditions that he
knew would lead to such analytical solutions. It is unlikely that one may be
able to sum the time series exactly for two arbitrary functions $u(x,0)$ and
$u_{t}(x,0)$; therefore, in an actual physical problem one has to rely on
the accuracy of truncated time series that most probably will provide a
quite poor description of the dynamics of the problem beyond a relatively
small neighborhood of $t=0$.

Obviously, we may substitute Eq.~(\ref{eq:u_t_series}) into Eq.~(\ref
{eq:B(m,n)}) and obtain the coefficients $f_{j}(x)$ in a straightforward
way. However, Y\i ld\i r\i m preferred the application of the lengthier HPM.
To make this point clearer let us write Eq.~(\ref{eq:B(m,n)}) as $%
u_{tt}=F(u,u^{\prime })$, where $u^{\prime }$ collects all the derivatives
of $u(x,t)$ with respect to $x$, and $F$ is a nonlinear function. Taking
into account that $F(u,u^{\prime })$ can be expanded in a Taylor series $%
F=F_{0}+F_{1}t+\ldots $ we easily derive a recurrence relation for the
coefficients $f_{j}$:
\begin{equation}
f_{j+2}=\frac{F_{j}}{(j+1)(j+2)},\,j=0,1,\ldots  \label{eq:fj}
\end{equation}
where, obviously, $F_{j}$ depends on $f_{i}$, $i=0,1,\ldots ,j$. Since we
know $f_{0}$ and $f_{1}$ as already indicated above then we obtain all the
other coefficients $f_{j}(x)$ with $j>1$ in a straightforward way.
Considering that this procedure is a textbook example of calculus, is there
any necessity for the application of HPM or any other more or less elaborate
method to obtain the same series?. May the reader answer this question by
him/herself.

Y\i ld\i r\i m obtained a kind of travelling--wave solutions to the $B(m,n)$
equations and stated that ``exact solutions with solitary patterns are of
important significance''. Y\i ld\i r\i m also identified a kind of
superposition of waves that is unexpected in nonlinear problems: ``This
paper reveals first time that in some special cases, addition of two
solitary solutions satisfies exactly the nonlinear equation. Such phenomenon
requires further mathematical study and physical explanation.'' In what
follows we analyze Y\i ld\i r\i m results as well as such phenomenon.

Following the standard procedure for obtaining travelling waves we write
\begin{equation}
u(x,t)=U(\xi ),\,\xi =x\pm vt  \label{eq:U(xi)}
\end{equation}
so that Eq.~(\ref{eq:B(m,n)}) becomes
\begin{equation}
\left[ v^{2}U+U^{m}-(U^{n})^{\prime \prime }\right] ^{\prime \prime }=0
\label{eq:U_diffeq}
\end{equation}
where the prime indicates differentiation with respect to $\xi$. For present
purposes it is sufficient to consider polynomial solutions of the form
\begin{equation}
U(\xi )=\sum_{j=-M}^{N}a_{j}e^{\alpha j\xi }  \label{eq:U_trial}
\end{equation}
keeping in mind that other expressions (like, for example, rational
functions) may also be suitable.

We first discuss the $B(2,2)$ equation
\begin{equation}
u_{tt}+(u^{2})_{xx}-(u^{2})_{xxx}=0  \label{eq:B(2,2)}
\end{equation}
and purposely leave the boundary conditions unspecified. In this particular
case we have
\begin{equation}
U(\xi )=a_{2}e^{\xi /2}+a_{-2}e^{-\xi /2}-\frac{2}{3}v^{2}
\label{eq:U_B(2,2)}
\end{equation}
where $\alpha =1/4$ and $a_{2}$ and $a_{-2}$ are arbitrary coefficients.

By means of HPM Y\i ld\i r\i m\cite{Y09} derived two solutions for Eq.~(\ref
{eq:B(2,2)}) with different boundary conditions and then combined them into
a more general one of the form
\begin{eqnarray}
u^{Y}(x,t) &=&a\cosh ^{2}\left( \frac{\xi }{4}\right) +b\sinh ^{2}\left(
\frac{\xi }{4}\right) ,  \nonumber \\
a &=&-\frac{4}{3}Lv^{2},  \nonumber \\
b &=&\frac{4}{3}(1-L)v^{2}  \label{eq:u_Y}
\end{eqnarray}
Y\i ld\i r\i m\cite{Y09} obtained other solutions by addition of a constant
to the arguments of the hyperbolic functions. This strategy for generating
new solutions is successful because if $U(\xi )$ is a solution to Eq.~(\ref
{eq:U_diffeq}), then $U(\xi +\xi _{0})$ is also a solution (with, of course,
different boundary conditions). Eq.~(\ref{eq:u_Y}) is a particular case of
our Eq.~(\ref{eq:U_B(2,2)}) with $a_{2}=a_{-2}$ as follows from the fact
that
\begin{equation}
u^{Y}(x,t)=\frac{a+b}{4}(e^{\xi /2}+e^{-\xi /2})-\frac{2}{3}v^{2}
\label{eq:u_Y_2}
\end{equation}
At this point it is worth noticing that the solutions that we are discussing
are unbounded and, therefore, of little or no physical utility, whatsoever.
However, there seems to be some interest in them anyway\cite{Y09}.

As promised above we now address the Y\i ld\i r\i m's superposition
principle that is embodied in the solution~(\ref{eq:u_Y}) to Eq.~(\ref
{eq:B(2,2)}). Notice that $u^{Y}(x,0)$ and $u_{t}^{Y}(x,0)$ depend on the
coefficients of the linear combination (through $L$). Therefore, the
solutions with, say, $L=0$ and $L=1$ exhibit different boundary conditions
and thereby correspond to different physical problems. In other words, Y\i
ld\i r\i m's superposition principle is a linear combination of solutions to
different problems. We do not think that such a property is of any physical
relevance but, of course, we may be wrong.

In the case of the $B(3,3)$ equation
\begin{equation}
u_{tt}+(u^{3})_{xx}-(u^{3})_{xxx}=0  \label{eq:B(3,3)}
\end{equation}
we have
\begin{eqnarray}
U(\xi ) &=&a_{3}e^{\xi /3}+a_{-3}e^{-\xi /3}  \nonumber \\
\alpha &=&\frac{1}{9}  \nonumber \\
a_{3}a_{-3} &=&-\frac{3}{8}v^{2}  \label{eq:U_B(3,3)}
\end{eqnarray}
that leads to the two solutions obtained by Y\i ld\i r\i m when $%
a_{3}=-a_{-3}=\pm \sqrt{6}v/4$.

Finally, we mention that equations (\ref{eq:B(m,n)_0}) and (\ref{eq:B(m,n)})
are related by the transformation $t\rightarrow it$ (or $v\rightarrow iv$
for the travelling waves). Therefore, we easily derive the solutions to Eq.~(%
\ref{eq:B(m,n)_0}) from the ones just obtained.

We think that it is unnecessary to consider other cases because equations (%
\ref{eq:B(m,n)_0}) and (\ref{eq:B(m,n)}) do not appear to have any physical
application whatsoever. They are merely tailor--made toy problems for the
application of approximate methods of doubtful utility. Besides, the results
and discussion above are more than enough for making our point.

Summarizing:

\begin{itemize}
\item  The HPM proposed by Y\i ld\i r\i m\cite{Y09} to treat
Boussinesq--like equations simply leads to the Taylor series of the
travelling--wave solutions about $t=0$; consequently, we can obtain the same
results more easily by means of the latter well known approach. Obviously,
we may apply the Taylor--series approach with initial conditions different
to those that lead to travelling waves, but in such more general cases we
are not certain to obtain analytical solutions.

\item  Instead of resorting to the HPM, one can obtain more general
solutions by means of the textbook method of travelling--waves reduction.

\item  Y\i ld\i r\i m's superposition principle is just the linear
combination of solutions to models with different boundary conditions and,
therefore, it is of questionable physical significance.

\item  Finally, we stress the fact that many applications of approaches like
HPM (see below) have been restricted to tailor--made toy problems with exact
analytical solutions. As in the case analysed here the authors commonly
obtain Taylor series in a rather indirect but more fashionable way.
\end{itemize}

In earlier reports we have argued that some new applications of variational
and perturbation approaches (VAPA) like HPM, homotopy analysis method (HAM)
(quite similar to HPM), variation--iteration method (VIM) and Adomian
decomposition method (ADM) are responsible for some of the poorest
scientific papers ever published in supposedly respectable journals\cite
{F07,F08b,F08c,F08d,F08e,F08f,F09}. There seems to be a great interest in a
new physics, or mathematical physics, based on Taylor expansions of
nonlinear problems, solutions to the Schr\"{o}dinger equation that are not
square integrable, tortuous ways for obtaining well--known analytical
results, laughable treatments of chemical reactions, etc\cite
{F07,F08b,F08c,F08d,F08e,F08f,F09,F09c}. The most remarkable example is
provided by a predator--prey model that predicts a negative number of rabbits%
\cite{F08d}. The VAPA journals that publish such kind of $\mathit{scientific}
$ articles do not accept criticisms or comments on them. JMP has engrossed
the distinguished list of VAPA journals.

Of course we may have a completely wrong point of view. For that reason in
what follows we show the critical opinion of a prestigious anonymous referee:

\textsl{Referee \#1 (Comments to the Author):}

\textsl{This paper is a comment on Yildirim's paper, but the title is too
large. }

\textsl{Yildirim's might have some demerits in one way or another, the
homotopy perturbation method does work for the Boussinesq-like equation. }

\textsl{If the author wants to write a comment on the homotopy perturbation
method, the author should base on Ji-Huan He's publications on HMP. }

\textsl{The method is an asymptotic method , though the method can also lead
to a convergent series solution. }

\textsl{An elementary introduction to the method is given in the following
editorial article: He JH. Recent development of the homotopy perturbation
method , TOPOLOGICAL METHODS IN NONLINEAR ANALYSIS 31(2008) 205-209 }

\textsl{The author suggests a straightforward method , that we can submit
Eq.(4) into (2), and same result can be obtained . This is true , and such
method becomes invalid for most nonlinear equations. }

\textsl{The author ignors the connection between his method and famous known
method of separation of variables used by Joseph Fourier in 1822. }

\textsl{The author further assumes that the solution can be expressed in (8)
with exp-polynomials, again this special suggestion works, but it does work
for most nonlinear problems, the author should follow the exp-function
method in this case. }

\textsl{The author gives a solution without considering the boundary
conditions, this has only mathematical interesting, and has no physical
meanings at all. }

\textsl{Such discuss should be directly addressed to DR.Yildirim, not
suitable for publication. }

\textsl{I note that the author has already posted his comments on arXiv,
this is enough.}

This is an interesting example of the quality of the JMP refereeing
service. We should be grateful to be enlightened this way.

\end{document}